\documentclass[aps,pra,nofootinbib,preprint,amsmath,amssymb,floatfix]{revtex4}
\usepackage{color}
\usepackage{graphicx}

\begin{document}

\newcommand{\tc}{\textcolor}
\newcommand{\g}{blue}
\newcommand{\ve}{\varepsilon}
\title{Remarks on  the Abraham-Minkowski problem, from the formal and from the experimental side}         

\author{ Iver Brevik$^1$, Masud Chaichian$^2$, and Ion I. Cot\u{a}escu$^3$  }      
\affiliation{$^1$Department of Energy and Process Engineering, Norwegian University of Science and Technology,  N-7491 Trondheim, Norway}

\affiliation{$^2$ University of Helsinki and Helsinki Institute of Physics, FIN-00014 Helsinki, Finland }
\date{\today}          

\affiliation{$^3$ West University of Timisoara, V. Parvan Ave. 4, 300223 Timisoara, Romania}

\begin{abstract}
We analyze the Abraham-Minkowski problem known from  classical electrodynamics from two  different perspectives. First, we follow a formal approach, implying use of manifolds with curved space sections in accordance with Fermat's principle, emphasizing that the resulting covariant and contravariant components of the photon four-momentum is a property linked to the {\it Minkowski} theory only. There is thus no link to the Abraham theory in that context. Next we turn to the experimental side, giving a brief account  of older and newer  radiation pressure experiments that clearly show how the Minkowski photon momentum is preferable under optical conditions. Under low-frequency conditions, where experimental detection of the individual oscillations predicted by the Abraham term are possible, the picture is however quite different.
PACS: 04.20.Cv
\end{abstract}

\maketitle

\bigskip

\section { Introduction}

A  historical controversy concerns the Abraham \cite{A}, versus  the Minkowski \cite{M}, definitions of the momentum of a photon  propagating  in a continuous medium (some references are \cite{1,2,3,4,5,6,7,8,8a,8b}). In  classical electrodynamics the Minkowski momentum density is ${\bf g}^{\rm M}= {\bf D\times B}$, whereas the Abraham momentum density is ${\bf g}^{\rm A}={\bf E\times H}/c^2$. The Planck principle of inertia of energy, saying that the relationship between momentum density $\bf g$ and the Poynting vector $ \bf S$ is ${\bf g}={\bf S}/c^2$, is thus broken in the Minkowski case.

The phenomenological theory can be quantized if the underlying electromagnetic energy-momentum tensor is such that it permits the construction of the Fock space. In practice, this will be possible only for the Minkowski case, since then the energy-momentum tensor is divergence-free, thus securing four-vector properties of the energy and momentum components  of the radiation field (cf., for instance, Refs.~\cite{brevik17} and \cite{brevik20}.

The Abraham-Minkowski problem has many facets, and we will here focus on two of them. First we will approach the problem from a formal side,
related to the wave-particle duality. As formulated by the  Barnett \cite{Bar}, the Abraham and Minkowski momenta are respectively the kinetic and canonical optical momenta. The formalism from general relativity can be used to illustrate these formal properties, by considering the optical medium as a pseudo-Riemannian manifold where the photon momenta  satisfy the standard null condition. We assume for simplicity a transparent and  isotropic medium,  at rest only. Thus, the photon momenta ${\bf p}^{\rm A}$ and ${\bf p}^{\rm M}$ are related via
\begin{equation}\label{OP}
E=|{\bf  p}^{\rm A}| nc=|{\bf p}^{\rm M}|\frac{c}{n}\,
\end{equation}
where $E$ is the photon energy.

Here we should emphasize that these formal properties do {\it not} reflect the physical difference between the Minkowski and Abraham theories. The quite elegant formalism relates to the four-vector property of photon momentum, a property possessed by the Minkowski theory but not by the Abraham one. This point has some times led to misunderstandings.

We will in the next item 2  consider  photons propagating  in  a $(1+3)$-dimensional  local-Minkowskian curved manifold $M$  in which  the propagation occurs  along  null geodesics while the electromagnetic waves are determined  by the Maxwell equations without sources.  Omitting gravity,  these requirements are satisfied by a pseudo-Riemannian metric (a Fermat manifold) resulting from the   invariant Fermat principle applied to the proper time instead of the optical length as in the general relativistic case \cite{Ob,9,10,11,12}.

We then move on to physical considerations, presenting a brief review over optical experiments and singling out those that actually show the importance of the Minkowski photon momentum and those that do not.  Actually, most of the experiments are not important in this respect, as they show only the action of electromagnetic forces at dielectric boundaries. Forces of that type are equally well described by the Abraham tensor as by the Minkowski tensor, and have little to do with photon momentum in matter.

Finally, as a contrast, we focus attention on how the theory becomes modified if the frequencies are so low that the fluctuations arising from the so-called Abraham term become detectable directly.

\section {Fermat manifolds}

 Let $M$ denote  a transparent isotropic ideal static medium at rest in which the light is neither absorbed nor dispersed.  We consider the static metric $g$ of components $g_{00}=-1\,, \quad g_{ij}(x)=\delta_{ij}n(x)^2\,, \quad i,j,k,...=1,2,3\,$
where $n(x)$ is the refractive index.   Introduce   Cartesian coordinates, $x^{\mu}$, and write the line element as
\begin{equation}\label{ds}
ds^2=g_{\mu\nu}(x)dx^{\mu}dx^{\nu}=-c^2 dt^2+n(x)^2\left[ (dx^1)^2+(dx^2)^2+ (dx^3)^2\right]\,,
\end{equation}
where  $n(x)$ depends only on the space coordinates $x^i$.

In this geometry  test particles move along geodesics $x^i=x^i(t)$, with  velocity $\dot{x}^i=\frac{dx^i}{dt}=v^i$, determined  from  the geometric variational principle
\begin{equation}\label{prin}
\delta S=\delta \int_{t_1}^{t_2} \sqrt{-ds^2}=\delta \int_{t_1}^{t_2}\sqrt{c^2-n^2\left[ (\dot{x}^1)^2+ (\dot{x}^2)^2+ (\dot{x}^3)^2\right]}\,dt=0.
\end{equation}
This can be looked upon  as the relativistically invariant  Fermat's  principle, applied to  proper time instead of to optical path  \cite{12}.  The photons propagate in $M$ along null geodesics whose equations of motion are
\begin{equation}
\dot{v}^k={\vec{v}\,}^2\,\frac{1}{n}\,\partial_k n  - 2\frac{v^k}{n} (v^i\partial_i n)\,, \quad  k=1,2,3\,,
\end{equation}
allow a prime integral that can be set as $n^2 {\vec{v}\,}^2=c^2$ corresponding to the null condition $ds^2=0$. Thus we recognize the simple relation $v=\frac{c}{n}$ that holds for isotropic media.

Now, we introduce  the dual  (or contravariant) metric tensor $\bar{g}$, with  components
\begin{equation}\label{gu}
g^{\mu\nu}={\rm diag}\left(-1, \frac{1}{n^2},\frac{1}{n^2},\frac{1}{n^2}\right).
\end{equation}
They appear  in the Maxwell equations
$\nabla_{\mu}F^{\mu\nu}= 0$,
with $g=|\det(g)|$. The   Lorentz condition is $\nabla_{\mu}A^\mu=0\,.$
Combining these equations one obtains the wave equation for  the electromagnetic potential $A_{\mu}$.

On the manifold $M$ we may introduce orthogonal non-holonomic local frames defined by the tetrad fields $e_{\hat\alpha}=e_{\hat\alpha}^{\mu}\partial_{\mu}$ and the corresponding  co-frames given by the 1-forms $\omega^{\hat\alpha}=\omega^{\hat\alpha}_{\mu}dx^{\mu}$,  labeled by local indices $\hat\alpha,...\hat\mu,\hat\nu...=0,1,2,3$. These frames are orthogonal and dual to each other satisfying the duality condition $\omega^{\hat\alpha}(e_{\hat\beta})=\delta^{\hat\alpha}_{\hat\beta}$ and the orthogonality condition $g(e_{\hat\alpha}, e_{\hat\beta})=\eta_{\hat\alpha\hat\beta}$.   It is convenient to choose the diagonal tetrad gauge,
\begin{eqnarray}
e_{\hat 0}=\partial_t \,,&\quad& e_{\hat i}=\frac{1}{n}\partial_i \,,\label{e}\\
\omega^{\hat 0}=c dt\,, &\quad &\omega^{\hat i}=n dx^i   \,,\label{o}
\end{eqnarray}
since in this gauge the space axes of the local frame are parallel with those of the space coordinates. However,  the normalization is different such that the line element (\ref{ds}) can be written as $ds^2=\eta_{\hat\alpha\hat\beta}\omega^{\hat\alpha}\omega^{\hat\beta}$ where
$\eta={\rm diag}(-1,1,1,1)$ is the Minkowski metric of the flat model of $M$. In our convention   the natural indices are raised or lowered by the metric tensor $g$  while for the local indices we have to use the metric $\eta$.

\noindent {\it  On the photon four-momentum.- }
Photon momentum is defined by the de Broglie relations that hold in static and homogeneous media where the wave vector is well-defined. For a homogeneous medium  $n$ is independent of the space coordinates and so   the Fermat manifold remains flat.
In  such a  manifold the photon moves along  linear null geodesics with constant velocity satisfying the null condition. In the Coulomb gauge, with $A_0=0$,  the wave equation  becomes
\begin{equation}
\square A_i=\Delta A_i -\frac{n^2}{c^2}\,\partial_t^2A_i=0\,,
\end{equation}
corresponding to the speed $c'=c/n <c$ in $M$.

The wave equation  has plane-wave solutions of the form $
A_i(t,\vec{x})\propto \varepsilon_i e^{ik_{\mu}x^{\mu}}$,
where $\varepsilon $ is the polarization vector while $k_\mu$ is the wave four-vector. Then, according to the de Broglie formulas, we may define the covariant  components $p_\mu$ of the photon four-vector,
\begin{equation}
p_{\mu}=\hbar k_{\mu}=\left(-\frac{E}{c},p_1,p_2,p_3 \right)\,.
\end{equation}
Writing the   null condition in the form
 \begin{equation}\label{NM}
0=g^{\mu\nu}p_{\mu}p_{\nu}= \frac{1}{n^2}p_ip_i-\frac{E^2}{c^2},           
\end{equation}
it is natural to identify  $p_i$ with the spatial components of the Minkowski photon four-momentum, $p^{\rm M } \equiv  |{\bf p}^{\rm M}|=    nE/c$. It agrees with  the second member of Eq.~(\ref{OP}).

We may define also  the contravariant components of the  four-momentum,
\begin{equation}\label{NA}
p^{\mu}=g^{\mu\nu}p_{\mu}=\left(\frac{E}{c},\frac{p_1}{n^2},\frac{p_2}{n^2},\frac{p_3}{n^2} \right)=\left(\frac{E}{c},p^1,p^2,p^3 \right)\,,
\end{equation}
and write the null condition in the form
\begin{equation}
0=g_{\mu\nu}p^{\mu}p^{\nu}=n^2 p^i p^i  - \frac{E^2}{c^2}.
\end{equation}

In addition, thanks to our approach we can define the {\em local} momentum $\hat p$  in the orthogonal frames defined by the tetrad gauge (\ref{e}) and (\ref{o}). According to the general
definitions, $\hat p^{\hat\alpha}=\omega^{\hat\alpha}_{\mu}p^{\mu}$ and  $\hat p_{\hat\alpha}=e_{\hat\alpha}^{\mu}p_{\mu}$, we find the components
\begin{equation}
\hat p^{0}=-\hat p_{0}=\frac{E}{c}\,,\quad \hat p^i=\hat p_i=n p^i =\frac{p_i}{n} \,,
\end{equation}
which satisfy the identity
\begin{equation}
0=\eta_{\hat\alpha\hat\beta}\hat p^{\hat\alpha}p^{\hat\beta}=(\hat p^1)^2+(\hat p^2)^2 +(\hat p^3)^2-\frac{E^2}{c^2}\,.
\end{equation}

The conclusion is that in our manifold $M$  the Abraham-type momentum is the vector field
\begin{equation}\label{pA}
p_A=p^{\mu}\partial_{\mu}=\hat p ^{\hat\alpha}e_{\hat\alpha} \in TM\,,
\end{equation}
while that of Minkowski-type the dual co-vector (or 1-form)
\begin{equation}\label{pM}
p_M=p_{\mu}dx^{\mu}=\hat p_{\hat\alpha}\omega^{\hat\alpha}\in T^*M\,.
\end{equation}
The notations $TM$ and $T^*M$ stand for the tangent  and respectively cotangent bundles on $M$ \cite{Geo}. Now the null conditions (\ref{NM}) and (\ref{NA}) as well as their particular cases (\ref{OP}) are given just by the duality relation
\begin{equation}
p_M(p_A)=g(p_A,p_A)=\bar{g}(p_M,p_M)=0\,.
\end{equation}
Thus we defined coherently the momenta  $p_A$ and $p_M$ in flat Fermat manifolds solving at least {\em  formally} the Abraham-Minkowski dilemma.

However, this formal conciliation enters in collision with the actual electrodynamics of continuous media which traditionally cannot be decoupled to special relativity where  $p_\mu$ and $p^\mu$  are the covariant and contravariant components of one and the same Minkowski four-vector. Nevertheless, if we accept as an hypothesis a future non-Minkowskian electrodynamics, in which some media have Fermat geometry,  then the above presented formalism could explain why  in  the Abraham formulation the radiation energy and momentum {\it  do not make up a four-vector at all} in the sense of special relativity. The reason for this is that the Abraham energy-momentum tensor is not divergence-free, even in a homogeneous medium. Note that this was pointed out by M{\o}ller a long time ago \cite{moller72}, and has been emphasized by one of us several authors several times \cite{2,brevikMPLA,brevikPRA} also.

\section{ On  experiments}

As many nice radiation pressure experiments have been reported in the literature it  becomes important to check if one, or more, of the theoretical explanations of them stands out as the most favorable one. And this actually turns out to be the case. As mentioned above, all optical experiments of this sort that we are aware of, become explainable by the Minkowski theory in a simple and straightforward way. A nontrivial point, however, is whether a specific  experiment gives information about photon momentum, or if it merely shows the action of electromagnetic forces at dielectric boundaries. One of us gave recently an overview of this sort \cite{brevikMPLA}; we will not repeat that here, but will give a brief summarizing survey and moreover focus on a very new experiment that also seems to be valuable in this context.

We will start with the general expression for the electromagnetic force density $\bf f$ in an isotropic medium:
\begin{align}
{\bf f}=&\rho{\bf E}+{\bf J\times B}+\frac{1}{2}\varepsilon_0 {\bf \nabla}\left[ E^2\rho_m\frac{\partial \varepsilon}{\partial \rho_m}\right]+
\frac{1}{2}\mu_0 {\bf \nabla}\left[ H^2\rho_m\frac{\partial \mu}{\partial \rho_m}\right] \notag \\
&-\frac{1}{2}\varepsilon_0 E^2{\bf \nabla} \varepsilon -\frac{1}{2}\mu_0H^2{\bf \nabla}\mu +\frac{n^2-1}{c^2}\frac{\partial}{\partial t}{\bf (E\times H)}.
\end{align}
We assume no charges or currents, $\rho= {\bf J}=0$, assume the material to be nonmagnetic, and omit the two electrostriction and magnetostriction terms ($\rho_m$ is the matter density). We write the constitutive relations as ${\bf D}=\varepsilon \varepsilon_0 {\bf E}, \, {\bf B}=\mu_0{\bf H}$, so that $\varepsilon$ becomes nondimensional. The basic expression above thus reduces to the form
\begin{equation}
{\bf f}^{\rm A}= -\frac{1}{2}\varepsilon_0 E^2{\bf \nabla} n^2  +\frac{n^2-1}{c^2}\frac{\partial}{\partial t}{\bf (E\times H)}, \label{abraham}
\end{equation}
with $n^2=\varepsilon$. This is the Abraham force density. The force consists of two terms, the first acting at dielectric boundaries typically, the second (usually called the Abraham term) acting in the bulk. The second term is however rapidly fluctuating out in an optical wave.

The Minkowski force density ${\bf f}^{\rm M}$  is equal to the expression above, except from the Abraham term,
\begin{equation}
  {\bf f}^{\rm M}= -\frac{1}{2}\varepsilon_0 E^2{\bf \nabla} n^2. \label{minkowski}
  \end{equation}
In a homogeneous interior this force thus vanishes.

The action of the Abraham term is to produce an accompanying mechanical momentum density
\begin{equation}
{\bf g}_{\rm mech}= \frac{n^2-1}{c^2}{\bf E\times H},
\end{equation}
which together with the Abraham momentum density ${\bf g}^{\rm A}={\bf (E\times H)}/c^2$ gives rise to the total propagating Minkowski momentum density
\begin{equation}
{\bf g}^{\rm M}={\bf g}^{\rm A} + {\bf g}_{\rm mech}={\bf D\times B}.
\end{equation}
When considering  some concrete radiation pressure experiments, the central question becomes in each case: does the observed result show the action of the Abraham force (\ref{abraham}), or does it only show the force at the boundaries, i.e., the Minkowski force (\ref{minkowski})? We first list a few examples belonging to the same (actually the second) category:

A classic experiment is that of Ashkin and Dziedzic \cite{ashkin73}, in which a pulsed laser beam at 0.53 ${\mu}$m  was sent down on a water surface, resulting in a weak outward bulge ($0.9~\mu$m height) of the surface. The reason for the weakness was the high surface tension for water. Later experiments, operating with micellar liquids in the vicinity of the critical point, led to much larger surface elevations, about $70~\mu$m \cite{casner03}, because of the very significant reduction of surface tension in this case.   The recent experiment of Astrath {\it et al.} \cite{astrath14} is essentially of Ashkin-Dziedzic type.  Experiments on curved surfaces have also been done, by Zhang and Chang on water droplets \cite{zhang88}, and by Zhang {\it et al.} on mineral oil and water surfaces \cite{zhang15}. An interesting recent experiment of Kundu {\it et al.} \cite{kundu17} demonstrates how a horizontal graphene oxide plate becomes deflected when illuminated by a weak continuous laser beam from above.

All these experiments have one property in common: they are {\it explainable in terms of radiation surface forces on boundaries only} (although it is  sometimes claimed that they favor  the Abraham theory; cf., for instance, the critical note by one of us \cite{brevikPRA} on this point). They thus show the action of the force (\ref{minkowski}). In that way they do not have a direct bearing on the Abraham-Minkowski problem.

One may ask: which experiments are then able to show the action of the Abraham term? We may summarize a few cases of that kind as follows:

\bigskip

\noindent {\bf 1.  Radiation pressure on a metallic suspended mirror}

\noindent  The classic  experiments of Jones {\it et al.} \cite{jones54,jones78}, measuring the pressure on a metallic mirror  immersed in a dielectric liquid, still play a prominent role. The pressure was found to be proportional to the dielectric constant $n$ of the liquid, thus supporting the reality of the flowing Minkowski momentum density ${\bf g}^{\rm M}$.

\bigskip

\noindent {\bf 2. Photon drag in semiconductors}

\noindent The semiconductor experiment of Gibson {\it et al} \cite{gibson80} is also important in our context.  The longitudinal electric field produced by the momentum transfer from an incident radiation field was measured. This field was produced by charges driven down in the dielectric rod. The agreement with the Minkowski prediction for photon momentum was quite good  (this experiment was also commented upon in Ref.~\cite{brevik86}).

\bigskip

\noindent {\bf 3. Direct measurement of photon momentum}

\noindent  A third and noteworthy experiment is that of Campbell {\it et al.} \cite{campbell05},  making a direct measurement of the photon recoil in a Bose-Einstein condensate. The photon momentum was found to be $\hbar k = \hbar n\omega/c$, just as predicted in the Minkowski theory.

\bigskip

\noindent {\bf 4.  Photon drag in thin metal films}

\noindent  An interesting and very recent result is the measurement of  Minkowski's photon momentum, actually as a by-product, in the photon drag experiment of Strait {\it et al.} \cite{strait19}. The main purpose of this experiment was not to test Minkowski momentum but rather to focus attention on the possibility of a counter intuitive reversion of the direction of  the induced electron flow caused by the photons. As this experiments is apparently not well known in the radiation pressure community, we will go into some detail, focusing on the part of the experiment on main importance here.

   The setup was a  SiO$_2$ hemisphere resting on a thin, flat Au layer. Incident photons  coming from the ambient air region, inclined at an angle  $\theta$ with the vertical, propagated orthogonally through the hemispherical surface  and exerted subsequently  a drag on the electrons in the plate. The  electromagnetic energy reflected back from the spherical surface was  low and neglected to the actual order of accuracy.
   We will consider only the case of {\it s} polarization, for which the radiation force was found to act in the forward direction, what is one would intuitively expect. The thickness of the Au plate was 35 nm, the free electron density was $5.9\times 10^{22}~$cm$^{-3}$, and an optical  pulse train of wavelength 800 nm was used. The angle $\theta$ varied between 0 and $\pm 60^{\rm o}$.

   The same measurement was made in the absence of the semicylinder. Comparison of the in-plane transduction factors denoted by $\xi_g^s$ (glass-metal) and $\xi_f^s$ (free space)  showed that to their ratio was equal to 1.53. This is roughly the same as the refractive index 1.45 for SiO$_2$. As the experimenters concluded themselves, this strongly indicates that the {\it s} wave photonic voltage was proportional to the Minkowski momentum $n\hbar \omega/c$. The experiment is clearly related  to the photon drag experiment discussed above.

   The case of {\it p} polarization was more complicated, and might involve second order modification channels as discussed in Ref.~\cite{strait19}. Analog results to those above were then not found.

\bigskip
\noindent {\bf 5. The case of low frequencies}

\noindent Thus far, we have considered optical frequencies and the obvious usefulness of Minkowski's momentum under those circumstances. Quite a different picture arises if we turn to the case of low frequencies, for which the oscillations predicted by the Abraham force term in Eq.~(\ref{abraham})  become directly observable. The prominent example of this kind of experiment is that of the Walkers and Lahoz reported in Refs.~\cite{walker75} and \cite{walker75a}.  A dielectric high-permittivity disk (barium titanate, $\varepsilon = 3620$ was suspended in the gravitational field and acted as a torsional pendulum. The inner and outer disk surfaces were coated with aluminium, and a harmonic voltage of magnitude $V_0 \approx 260~$V of the same periodicity as the mechanical pendulum was applied across the surfaces. When a constant axial magnetic field was applied, there arouse an axial Abraham torque $N_z$ on the shell. With $\omega/2\pi \approx 0.4~$Hz, the predicted torque of magnitude $N_z \approx 4\times 10^{-10}~$Nm was found to agree with the observations to within $\pm 10\%$. This is a clear demonstration of the reality of the Abraham force, under these special conditions. We have actually emphasized the importance of this experiment before \cite{2}, but it still does not seem to be well known.

\section{Concluding remarks}

We have shed light on  the Abraham-Minkowski problem from two quite different perspectives. Our formal approach was related to the Gordon metric  $g_G$  which is also a Finslerian metric \cite{Gord}, defined in  classical relativistic electrodynamics \cite{Thom}. If the medium   is isotropic and at rest,  the Fermat and Gordon  metrics are related through the conformal transformation,  $g={n^2}\,g_G$. It  means that these are  geometrically non-equivalent even though they may produce similar null geodesics.

It is however to be borne in mind that  these formal developments, although instructive, are not able to  solve the physical Abraham-Minkowski problem. For that purpose, one should look at the experiments. We have listed a few experiments that we consider to be critical enough to test the Minkowski photon momentum. As mentioned, we emphasize that all experiments that we are aware of in optics, are most conveniently explained in terms of the Minkowski theory.

\subsection*{Acknowledgments}

We are very grateful to Merab Gogberashivili,
Friedrich  Hehl and   Yuri Obukhov
for several useful discussions at different stages of this work, and to Jared H. Strait for correspondence concerning Ref.~\cite{strait19}.

\end{document}